\documentclass[10pt]{article}
\usepackage{graphicx}
\usepackage{bm}
\usepackage[usenames]{color} 
\usepackage{ulem} 
\usepackage{amsmath,amssymb} 
\usepackage{subfigure}

\usepackage[labelfont=bf,labelsep=period,justification=raggedright]{caption}

\makeatletter
\renewcommand{\@biblabel}[1]{\quad#1.}
\makeatother

\date{}

\begin{document} 
\begin{flushleft}
{\Large
\textbf{Intervention-Based Stochastic Disease Eradication}
}
\\Lora Billings$^{1,\ast}$, 
Luis Mier-y-Teran-Romero$^{2,3}$, 
Brandon Lindley$^{2}$, 
Ira B. Schwartz$^{2}$

\bf{1} Department of Mathematical Sciences, Montclair State University, Montclair, NJ 07043\\
\bf{2} US Naval Research Laboratory, Code 6792,Nonlinear System Dynamics Section, Plasma Physics Division, Washington, DC 20375\\
\bf{3} Johns Hopkins Bloomberg School of Public Health, Department of International Health, Baltimore, MD, 21205
\\
$\ast$ E-mail: billingsl@mail.montclair.edu
\end{flushleft}
 
\section*{Abstract}
Disease control is of paramount importance in public health with infectious disease extinction as the ultimate goal. Although diseases may go extinct due to random loss of effective contacts where the infection is transmitted to new susceptible individuals, the time to extinction in the absence of control may be prohibitively long. Thus intervention controls, such as vaccination of susceptible individuals and/or treatment of infectives, are typically based on a deterministic schedule, such as periodically vaccinating susceptible children based on school calendars. In reality, however, such policies are administered as a random process, while still possessing a mean period. Here, we consider the effect of randomly distributed intervention as disease control on large finite populations. We show explicitly how intervention control, based on mean period and treatment fraction, modulates the average extinction times as a function of population size and rate of infection spread. In particular, our results show an exponential improvement in extinction times even though the controls are implemented using a random Poisson distribution. Finally, we discover those parameter regimes where random treatment yields an exponential improvement in extinction times over the application of strictly periodic intervention. The implication of our results is discussed in light of the availability of limited resources for control.
 


\section*{Introduction}

Understanding the processes underlying disease extinction is an important problem in epidemic prediction and control. Currently, total eradication of infectious disease is quite rare, but continues to be a major theme in public health. Temporary eradication, sometimes called fade out, tends to happen in local spatial regions, and may be followed by the reintroduction of the disease from other regions \cite{Anderson91,Grassly2005,finkenstadt2002stochastic}. In the case of diseases that possess co-circulating strains such as influenza \cite{Minayev2009}, or dengue fever which has up to four strains \cite{cummings}, extinction may occur in one or more strains while the others persist. Fade out or extinction may also occur within host, as is the case in Hepatitis C and HIV \cite{Perelson}. Infectious disease transmission is also conjectured to be responsible for certain species extinction \cite{smith2006evidence,ladeau2007west}. Recently, large scale amphibian species have had major declines in population, which have been linked with the spread of disease \cite{skerratt2007spread}.

One main reason that diseases go extinct is due to the stochasticity that is inherent to populations of finite size \cite{Bartlett1957}. As a disease evolves in large finite population, there is the possibility of insufficient transmission for it to stay endemic. Therefore, in finite time, the number of infectious individuals can go to zero and the disease dies out \cite{Allen2000,bartlett49,gar03}. Other mechanisms that enhance extinction include small populations and resource competition \cite{DeCastro2005}, as well as heterogeneity in host--vector models \cite{Lloyd2007}.

To properly model the random interactions occurring in populations, the study of disease extinction requires a stochastic modeling approach. As a result of the random interactions, time series analysis and epidemic models exhibit stochastic fluctuations \cite{Andersson,Earn,Bokler,Patz}. The fluctuations may act as an effective force that drives the disease to vanish \cite{sbd109}. This force is composed of non-Markovian fluctuations which may overcome the instability of the extinct state. The non-Markovian nature is due to the pattern of the noise necessary to drive the system out of equilibrium from the attracting state to the extinct state. It allows the escape from a quasi-stationary state by overcoming the force that keeps the system near the attractor. Therefore, if the fluctuations are weak on average, extinction is considered a rare event. 

We remark here that although escape has been considered for Langevin types of systems, the theory we present here is for discrete finite populations modeled as a master equation. In continuous systems, a rigorous theory of escape rates for systems driven by white Gaussian noise was developed by Freidlin and Wentzell \cite{Freidlin_book}. It was also found that the escape rates should display a number of universal features, including scaling behavior \cite{Dykman1979a,Dykman1980}, which has been confirmed by many experiments \cite{Aldridge2005,Stambaugh2006,Chan2007,Vijay2009}. More recently, it was shown that the state of the system is coupled to a deterministic model of the noise shape \cite{ForgostonBMB}. In this setting, the optimal path is an unstable object, but may be associated with the dynamical systems idea of having maximum sensitive dependence to initial conditions \cite{SchwartzJRS}.

Vaccination and treatment programs are common methods used to speed up the extinction of a disease in a population \cite{Nasell1999}. In this paper, we aim to quantify how treatment or vaccination increase average extinction rates. We focus on a class of diseases with no immune response. Models with no immunity are suitable for many bacterial infections, such as meningitis, plague, and venereal diseases, as well as certain protozoan illnesses, such as malaria and sleeping sickness \cite{Hethcote1976}. The mean field SIS model is typically analyzed for this scenario. If the disease is endemic, the disease state is attracting and the extinct, or disease free state, is unstable. 

Associated with the parameters for a particular disease is the basic reproduction number $R_0$, which defines on average how many new cases appear over one infectious period per infective \cite{Anderson91}. When $R_0 < 1$, the extinct state is attracting, but when $R_0>1$, the extinct state is unstable. The bifurcation point at which the endemic state appears is at $R_0$=1. For example, it was approximated from the epidemiological data from England and Wales that the serogroup C meningococcal disease had $R_0=1.36$, \cite{Trotter}. In Africa, some malaria $R_0$ estimates are close to one, but others can be as high as 3,000 \cite{Smith_malaria}. This variation is attributed to environmental temperature variations and mosquito biology \cite{Paaijmans}. Therefore, several groups have identified the applicability for methods to analyze extinction in finite populations near bifurcation points. (See the review in \cite{ovaskainen2010stochastic}.) In both basic SIS~\cite{dyscla08,sbdl09} and SIR~\cite{kammee08} models, the mean times to extinction were analyzed as a function of $R_0$ very close to one. The range of parameters here is assumed to model extremely slow disease propagation in large population limits. 

In general, little work has been done in analyzing stochastic models with {\it random} vaccination or treatment. Most vaccination schedules are designed as periodic, especially for childhood and seasonal diseases \cite{bolker1996impact}. Each intervention typically has a prescribed (deterministic) schedule, or distribution, but the extinction event is still random. In \cite{schwartz2004}, a yearly vaccination pulse that was used as a control parameter was shown to prevent large amplitude disease outbreaks in a modified stochastic seasonal SEIR model. More recent work has considered the effect of deterministically imposed transitions on the rates of extinction \cite{Dykman_vacc}. Specifically, \cite{Dykman_vacc} showed that a limited supply of vaccine can be optimally distributed to the susceptible population to change the rates of extinction exponentially in a range of low $R_0$. 

Thus, one of the main problems in understanding treatment and/or vaccination scheduling is that deterministic schedule models are {\it not} an accurate representation of the process. A more realistic scenario is that, on average, treatment scheduling has a mean period or cycle, but is itself a random process. In this paper, we study a randomly distributed treatment program of infected individuals. We are interested in evaluating treatment distributions by minimizing the mean time to extinction for the disease. Running simulations are computationally expensive and sensitive to population size. The theory presented here provides an alternate method to approximate the mean time to extinction. In our models, we identify conditions for which the escape rate theory applies and control strategies are effective. In particular, we derive explicit scaling functions of the exponent of the mean time to extinction in terms of $R_0$ and mean treatment levels. We also identify the most effective treatment schedules. Then, we compare the theory against numerical simulations for verification.

 
\section*{Methods}

The stochastic SIS compartmental model tracks the number of individuals in a population of size $N$ in one of two states: susceptible ($X_1$) or infected ($X_2$). In this model, we assume that the individuals become susceptible to the disease again upon recovery. The number of individuals in each state changes as birth, death, infection and recovery events occur. They are quantified by the following transition rates. New susceptible individuals are born at a rate $\mu$, and both susceptible and infected individuals die at the rate $\delta$. In this model, we assume that the individuals recover from the disease without significant mortality. To keep the population constant over time, on average, we assume that the birth rate is equal to the death rate, so $\delta = \mu$. If a susceptible comes in contact with an infected individual, the healthy individual may become infected. We use a mass action term with the contact rate $\beta$ to describe the flow of newly infected individuals from the susceptible group. We assume infected individuals recover at rate $\kappa$ and immediately re-enter the susceptible group. In addition, we assume a treatment schedule that occurs at randomly chosen times with a frequency $\nu$ times per year. Each time the treatment is applied, a fraction $g$ of all infected individuals recover and flow back into the susceptible class. This assumes the treatment has 100\% efficacy. To study the effect of treatments that are not as effective, a prefactor for $g$ could be added to capture the smaller efficacy. That case is similar to studying a smaller value for $g$, which is included in the parameter range $0 \le g < 1$ and therefore we do not study this issue separately. 

We now form the master equation that describes the time evolution of the stochastic system. The general theory of applying the WKB method to finite populations begins by assuming that the population of $N$ individuals may be divided into $n$ compartments. The number of people within each compartment is described by the $n$-dimensional state vector ${\bf X}=(X_1,X_2,\ldots,X_n)$ with integer components. Let the random state transitions governing the dynamics be described by the transition rates $W({\bf X},{\bf r})$, with ${\bf r}$ representing the increment in the change of ${\bf X}$, and ${\bf r}/N \ll 1$. Also, let the probability of finding the system in state ${\bf X}$ at time $t$ be $\rho ({\bf X},t)$. The general form of the master equation is
\begin{equation} \label{eq:master} 
\frac{\partial \rho ({\bf X},t)}{\partial t} = \sum_{\bf r} [W({\bf X}-{\bf r};{\bf r}) \rho({\bf X}-{\bf r},t) - W({\bf X};{\bf r}) \rho({\bf X},t)]. 
\end{equation} 
We assume that the system possess a single, strictly stationary solution, $\frac{\partial \rho}{\partial t}=0$, that corresponds to the extinct state, where one or more of the $n$ components of the vector $\bf{X}$ are equal to zero. 

In the large population limit (zero fluctuations), the mean field equations 
\begin{equation} \label{eq:det} 
\dot{\bf X} = \sum_{\bf r} {\bf r} \, W({\bf X};{\bf r}) 
\end{equation} 
describe the time evolution of the system's mean values in a deterministic manner. 

The next step is to find a probability distribution, $\rho({\bf X},t)$, that solves the master equation. When the probability current at the extinct state is sufficiently small, there will exist a quasi-stationary probability distribution with a non-zero number of infected individuals that decays into the stationary solution over exponentially long times. The rate at which the extinction of infected individuals occurs may be calculated from the tail of the quasi-stationary distribution. It has been shown that a WKB approximation to the quasi-stationary distribution allows one to approximate the mean-time to extinction with high accuracy for $N$ sufficiently large \cite{Black2011,dyscla08,sbdl09}. This approximation consists in assuming that the desired distribution has, to leading order, the exponential form 
\begin{equation}\label{eq:rho} 
 \rho({\bf x},t) = A \exp (-N {\cal S}({\bf x},t)), 
\end{equation} 
where ${\bf x} = {\bf X}/N$ is the normalized state (e.g., in an epidemic model, the fraction of the population in the various compartments) and $A$ is the normalization constant. The function $\mathcal{S}$ is given by an expansion of the form $\mathcal{S} = S + S_1/N + \ldots$.

Defining new scaled transition rates as $w({\bf x};{\bf r}) = W(N{\bf x};{\bf r}) /N $, inserting the ansatz of Eq.~\eqref{eq:rho} into the master equation, Eq.~\eqref{eq:master}, and keeping the first term only in the expansion for $\mathcal{S}$ yields a partial differential equation of the Hamilton-Jacobi form: 
\begin{equation}\label{eq:hamilton-jacobi} 
\frac{\partial {\cal S}}{\partial t} + H \left({\bf x} ,\frac{\partial {\cal S}}{\partial {\bf x} } \right) = 0. 
\end{equation} 
In analogy to Hamiltonian mechanics, the functions $H$ and ${\cal S}$ are called the Hamiltonian and the action, respectively. The derivative $\frac{\partial{\cal S}}{\partial {\bf x}}$ is called the momentum conjugate to ${\bf x}$ and is denoted by ${\bf p}$. With these approximations, the Hamiltonian function is 
\begin{equation} 
H({\bf x},{\bf p}) = \sum_{\bf r} w({\bf x};{\bf r}) [\exp ({\bf p} \cdot {\bf r}) -1] , 
\label{Hamiltonian} 
\end{equation} 
and the characteristic equations for Eq.~\eqref{eq:hamilton-jacobi} are precisely Hamilton's equations: 
\begin{equation} 
\dot{{\bf x}}=\partial_{{\bf p}}H({\bf x,p}), ~~~ 
\dot{{\bf p}}=-\partial_{{\bf x}}H({\bf x,p}). 
\label{eq:Hamilt_equations} 
\end{equation} 
For a more detailed description of the WKB method and other applications, see \cite{gang87,Dykman1994d,dyscla08}.

\subsection*{Model 1: Constrained SIS model with treatment}
 
In the first model, we make the following approximation in the SIS model to reduce the dimension of the problem. Assume the average population size is $N$ and constrain the population size such that $X_1+X_2 = N$. Therefore, we can consider the dynamics of the constrained SIS model in terms of infected individuals, $X_2$. Therefore, we need only to consider the following transition rates, which describe how individuals enter and leave the infected state: 
\begin{equation} \nonumber
\begin{array}{rll} 
W\bigl(X_2; 1\bigr) &=~\beta X_2(N-X_2)/N, & \mbox{new infections}; \\ 
W\bigl(X_2; -1\bigr) &=~(\mu+\kappa) X_2, & \mbox{recovery and natural death}; \\
W\bigl(X_2;-g X_2 \bigr) &=~\nu, & \mbox{treatment}.
\end{array} \label{trans:SIS} 
\end{equation} 
Therefore, the master equation from Eq.~(\ref{eq:master}) for the constrained SIS stochastic process is 
\begin{equation} 
\begin{array}{rl} 
\frac{\partial\rho(X_2,t)}{\partial t} = &(\mu+\gamma) \Big((X_2+1) \rho(X_2+1,t)-X_2 \rho(X_2,t)\Big) 
+ \nu\Big(\rho(X_2+g X_2,t)-\rho(X_2,t)\Big) \\ 
 &+\frac{\beta}{N}\Big((X_2-1)(N-(X_2-1)) \rho(X_2-1,t)-X_2 (N-X_2) \rho(X_2,t)\Big). 
\end{array} \label{master:1DSIS} 
\end{equation} 
Since the population variable in the master equation is integer-valued, the first argument of $\rho(X_2+gX_2,t)$ must be rounded to an integer in a consistent way. Since treatment is assumed to reduce the infectious number of individuals, in what follows, we choose to keep the integer part $[gX_2]$ of $gX_2$ (rounding down). Simplifying notation, we use $gX_2$ in the first argument of $\rho(X_2,t)$ as shorthand for $[gX_2]$. Note that for any particular realization of the master equation, the treatment ceases to have an effect whenever $gX_2<1$ (i.e., $X_2 < 1/g$) since for all those numbers of infecteds $[gX_2]=0$.

Following Eq.~\eqref{eq:det}, the associated deterministic model is
\begin{equation} 
\dot{X}_2 = \beta X_2(N-X_2)/N - (\mu+\kappa) X_2 -\nu g X_2.
\end{equation} 
This system has two steady states: the extinct state, $X_{2}=0$, and the endemic state, $X_2= N(1 - \frac{1}{R_0}- \frac{\nu g}{\beta})$, with $R_0=\beta/(\mu+\kappa)$. As expected, more treatment (increasing $g$ or $\nu$) decreases the number of infected individuals in the time-average. 

Equation \eqref{master:1DSIS} always possesses as a solution a stationary distribution where the probability of observing zero infected individuals $(X_2=0)$ is $\rho (0,t)=1$, which we identify as the extinct state. If $R_0 > 1$ and $N$ is large enough, Eq.~\eqref{master:1DSIS} will also possesses a quasi-stationary solution with an infected fraction fluctuating around an endemic state. Hence, if $R_0>1$ the disease can spread through a population and is considered endemic. 
 
To simplify the variables for our model, we normalize the population so $x_2=X_2/N$. Therefore, the Hamiltonian function for the SIS model is 
\begin{equation}
\label{eq:H SIS} 
H(x_2,p_2) = \beta x_2 (1-x_2) (e^{p_2}-1)+ (\mu+\kappa)x_2 (e^{-p_2}-1) +\frac{\nu}{N}(e^{-g N x_2 p_2}-1). 
\end{equation} 
The associated Hamiltonian system from Eq.~\eqref{eq:Hamilt_equations} is 
\begin{eqnarray} 
\begin{array}{l} 
\dot{x}_2 = \beta x_2 \left( 1-x_2 \right) {{e}^{p_2}}- (\mu+\kappa) x_2 {{ e}^{-p_2}}-\nu g x_2 e^{-g N x_2 p_2}, \\ 
 \dot{p}_2 = - \beta \left( 1-2\,x_2 \right) \left( {{ e}^{p_2}}-1 \right) - (\mu+\kappa) \left( {e}^{-p_2}-1 \right)+\nu g p_2 e^{-g N x_2 p_2} . 
\end{array} \label{SISHamEqns} 
 \end{eqnarray}
For the Hamiltionian system, the endemic steady state is $({x_2}_e,{p_2}_e)=(1 - \frac{1}{R_0}- \frac{\nu g}{\beta},0)$, while the stochastic die out state is $({x_2}_{0},{p_2}_{0})=(0,p^*)$, with $p^*$ implicitly defined by 
\begin{equation} \label{pstar}
\nu g \, p^* = \beta (e^{p^*} - 1) + (\mu + \kappa) (e^{-p^*} - 1) . 
\end{equation} 
Note that the endemic state exists only if ${x_2}_e > 0$. In addition, the endemic state has zero momentum, which is consistent with our expectation that the probability distribution have a maximum at $x_{2e}$ and hence $\frac{\partial S(x_{2e})}{\partial x_2} = p_{2e} = 0$. Since the variables $x_2$ and $p_2$ of the WKB approximation are not restricted to integer values, here the rounding of $gX_2$ poses no problem. However, this means that in the WKB framework, the treatment pulses have an effect at arbitrarily low values of $x_2$, in contrast to the master equation framework where, because of the rounding, treatment stops being applied whenever $X_2<1/g$. 


\subsection*{Model 2: Full SIS model with treatment} 

The second model is the unconstrained SIS treatment model in two-dimensions. We calculate $S$ and $I$ separately and allow the population fluctuation about $N$. If the fluctuations are small compared to $N$, the system will behave like the one-dimensional approximation. 

For the two-dimensional model, let the state vector be ${\bf X}=(X_{1},X_{2})$ and the transition vector be ${\bf r}=(r_{1},r_{2})$. The changes in the susceptible and infected populations for a single transition are represented by the transition rates: 
 \begin{equation} \nonumber
 \begin{array}{rll} 
 W\bigl({\bf X};(1,0)\bigr) &=~N \mu, & \mbox{birth of new susceptibles}; \\ 
 W\bigl({\bf X};(-1,0)\bigr)&=~\mu X_{1}, & \mbox{natural death for susceptibles}; \\ 
 W\bigl({\bf X};(0,-1)\bigr)&=~\mu X_{2}, & \mbox{natural death for infectious}; \\ 
 W\bigl({\bf X};(1,-1)\bigr)&=~\kappa X_{2}, & \mbox{natural recovery}; \\ 
 W\bigl({\bf X};(-1,1)\bigr)&=~\beta X_{1} X_{2}/N, & \mbox{new infections}; \\ 
 W\bigl({\bf X};(g X_2,-g X_2)\bigr)&=~\nu, & \mbox{treatment}. 
 \end{array} 
 \end{equation} 
Here, as in Model 1, the non-integer quantity $g X_2$ is rounded down to $[gX_2]$. We normalize the population so ${\bf x}=(x_{1},x_{2})$, with $x_1=X_1/N$ and $x_2=X_2/N$. Using the definition of the master equation, Eq.~\eqref{eq:master}, and the ansatz for the probability density in Eq.~\eqref{eq:rho}, the Hamiltonian from Eq.~\eqref{Hamiltonian} in normalized variables is 
\begin{equation} 
\begin{array}{ll} 
H(\bm{x},\bm{p}) = & \mu (e^{p_{1}}-1)+\beta x_{1}x_{2}(e^{-p_{1}+p_{2}}-1)+\kappa x_{2}(e^{p_{1}-p_{2}}-1) + \mu x_{1}(e^{-p_{1}}-1) \\ 
 & +\mu x_{2}(e^{-p_{2}}-1)+ \frac{\nu}{N} (e^{g x_2 N p_1-g x_2 N p_2}-1). 
\label{eq:SIR Hamiltonian} 
\end{array} 
\end{equation} 
The associated Hamiltonian system from Eq.~\eqref{eq:Hamilt_equations} is 
\begin{equation} \label{SIS2DHam}
\begin{array}{ll} 
\dot{x}_1 & = \mu e^{p_1} - \beta\, x_1 x_2 e^{-p_1 + p_2}+\kappa\, x_2\,{{ e}^{p_1- p_2}} 
 - \mu x_1 e^{-p_1} +\nu g q_2 e^{g x_2 N(p_1- p_2)} \\ 
\dot{x}_2 & = \beta\, x_1 x_2 e^{-p_1+ p_2} -\kappa x_2 e^{p_1- p_2} -\mu\, x_2 e^{-p_2} 
 -\nu g x_2 e^{g x_2 N(p_1- p_2) } \\ 
\dot{p}_1 & =-\beta\, x_2 \left( { e^{-p_1+ p_2}}-1 \right) - \mu \left( { e^{-p_1}}-1 \right) \\ 
\dot{p}_2 & =-\beta\, x_1 \left( e^{-p_1+ p_2} - 1 \right) - \kappa \left( e^{p_1- p_2} -1 \right) 
 -\mu \left( e^{- p_2} -1 \right) - \nu g \left(p_1- p_2 \right) e^{g x_2 N(p_1- p_2) } .
\end{array} 
\end{equation} 
Note once more that in the WKB framework, the treatment pulses have an effect for arbitrarily small $x_2$. For this Hamiltonian system, the endemic state is located at the point 
\begin{equation} 
(x_1,x_2,p_1,p_2)=\left(\frac{1}{R_0} + \frac{\nu g}{\beta},1 - \frac{1}{R_0}- \frac{\nu g}{\beta},0,0 \right) 
\end{equation} 
and the stochastic die out state is $(x_1,x_2,p_1,p_2)=\left(1,0,0,p^* \right),$ with $p^*$ defined implicitly as in Eq.~\eqref{pstar}.

\section*{Results} 

We now use these Hamiltonian models to approximate the mean time to extinction. Topologically, the solution that describes an extinction event in the Hamiltonian system will connect the endemic state (${\bf x}_a$) and stochastic die out state (${\bf x}_s$). The connecting manifold is, in fact, the most probable path to extinction when the stochastic system starts initially at the endemic state \cite{dyscla08,sbdl09}. This set of points is called the optimal path. Points on the path will also satisfy the Hamiltonian on the energy surface $H({\bm x} ,{\bm p})=0$ since it is a solution to the time-independent version of Hamilton-Jacobi equation, Eq.~\eqref{eq:Hamilt_equations}.

From the definition of the momentum, ${\bf p}(t)=\frac{\partial{\cal S}}{\partial {\bf x}}$, the action up to the zeroth order of $N$ along the optimal path can be found by evaluating 
\begin{equation} 
{\cal S}_{opt} = \int_{{\bf x}_a}^{{\bf x}_s} {\bf p}_{opt}(t) \cdot \dot{\bf x}_{opt}(t) \, dt .
\end{equation} 
Using this quantity, we approximate the mean time to extinction by 
\begin{equation}\label{eq:tau_ext} 
\tau_{ext} = B e^{N {\cal S}_{opt}} .
\end{equation} 
where $B$ is a prefactor that depends non-exponentially on the system parameters and on the population size. An accurate approximation of the mean-time to extinction depends on obtaining $B$ \cite{dykman_prefactor}. 

It is usually not a trivial task to identify the set of points that describe the optimal path. In some cases, it can be found analytically. One example is Model 1 with $g=0$, since Eq.~\eqref{eq:H SIS} has an explicit solution for $p_2$ when constrained to $H(x_2,p_2)=0$. An alternative approach is approximating the solution asymptotically. There are also several numerical approaches. One common method is to treat the system as a two point boundary value problem and solving using a shooting method \cite{keller1976}. In this paper, we use a generalized Newton's method that involves iterating an initial guess of the solution in the entire time domain \cite{BLindley}. Our initial guess must satisfy the property that the solution will stay asymptotically near the steady states except for a small, continuous, transition region between the two. This iterative procedure requires discretizing the model differential equations in time, using a second order approximation for the derivatives, and then solving the entire resulting system of nonlinear algebraic equations simultaneously. 

Equation \eqref{eq:tau_ext} holds if and only if a quasi-stationary distribution exists. This is the case if the time to extinction is exponentially long, i.e., $N {\cal S}_{opt} \gg 1$. Assuming that an endemic state does exist ($x_{2e} > 0$); $N {\cal S}_{opt} \gg 1$ will be satisfied for $N$ sufficiently large or, for fixed $N$, for an $R_0$ sufficiently large and $\nu g$ sufficiently small. The last conditions on the parameters mean that the disease should be highly transmissible and that the treatment should not be too intense. See the Supplementary Material, for a more detailed treatment on the necessary conditions for the quasi-stationary solution to exist.

\subsection*{Model 1}

Because the Hamiltonian system for the constrained model is in two dimensions, the first approximation to the action path simplifies to 
\begin{equation} 
{\cal S}_{opt} = \int^{0}_{{x_2}_e} p_2(x_2) \, d{x}_2 ,
\label{eq:1DTaction} 
\end{equation} 
with $p_2$ explicitly as a function of $x_2$, evaluating the integral along the optimal path. The Hamiltonian function of Eq.~\eqref{eq:H SIS} does not allow for an algebraic solution for $p_2(x_2)$ from the equation $H(x_2,p_2)=0$ that describes the path connecting the endemic state to the extinct state when $g\neq0$. Therefore, this integral in Eq.~\eqref{eq:1DTaction} must be approximated. 

For this model, an asymptotic approach can be used to approximate the action along the optimal path to extinction. We assume $g$ is small, which implies small treatment pulses. We expand $p_{2}$ in $g$ and substitute this expression into the equation $H(x_2,p_{2})=0$. The resulting expansion is 
\begin{align}\label{eq:path O(g)} 
p_{2}(x_2) = - \ln(R_0(1-x_2))\left(1-\frac{\nu g}{\beta(1-x_2)-(\mu+\kappa)}\right) + \mathcal{O}(g^2).
\end{align} 

\begin{figure}[!ht] 
\begin{center} 
\includegraphics[width=3.85in]{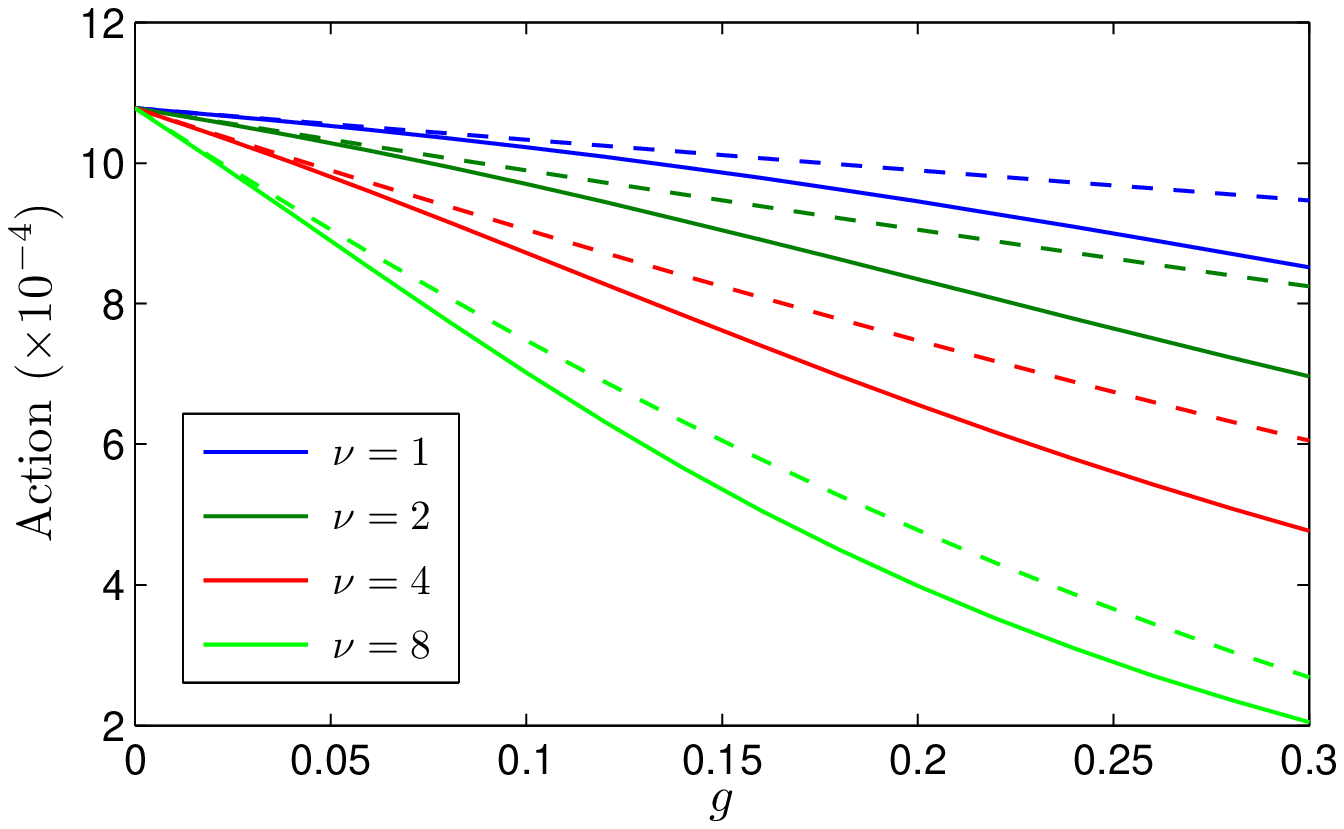} 
\end{center} 
\caption{
{\bf Comparing quantitative approximations of the action.} For Model 1,  plot of the numerical approximation of the action (dashed curve) and the asymptotic approximation (solid curve) as a function of the treatment, $g$. In this example, we use the parameters $\beta=105$ year$^{-1}$ and $N=8000$ people. As expected, the best agreement is for small $g$.
}
\label{fig:numasym} 
\end{figure}

The first term in the expansion $\mathcal{S} = S + S_1/N$ is given by Eq. \eqref{eq:1DTaction}. In \cite{Assaf2010}, the second term in the expansion of the action in powers of $N$ is given as more complicated integral along the path. We expand the two integrals giving $S$ and $S_1$ in powers of $g$ and evaluate them in closed form using computer algebra software. If we compare this asymptotic approximation to the numerical approximation for the action along the optimal path and evaluate Eq.~\eqref{eq:tau_ext}, we see excellent agreement as shown in the example in Figure \ref{fig:numasym}. In this example, we set the birth rate $\mu=0.2$ year$^{-1}$ and recovery rate $\kappa=100$ year$^{-1}$. For the remainder of the paper, we will use these parameters in examples for both models. These values provide results that can be clearly visualized and easily reproduced.

Note that while the action does not depend on the size of the population, to first order, the mean time to extinction does. The population size must be large enough to for the system to be quasi-stationary. Our model assumes that disease extinction is a rare event, which occurs in the tail of the distribution described by $\exp(-N{\cal S}_{opt})$. Conversely, the peak of the distribution occurs at the endemic state. As $R_0$ decreases to one, the distance between the endemic state and the disease free equilibrium decreases and the probability of the system having zero individuals in the infected state becomes significant. Therefore, the exponent must be large and negative, or equivalently the action must be sufficiently large compared to the population so that ${\cal S}_{opt} \gg 1/N$.

\begin{figure}[!ht] 
\begin{center} 
\includegraphics[width=4in]{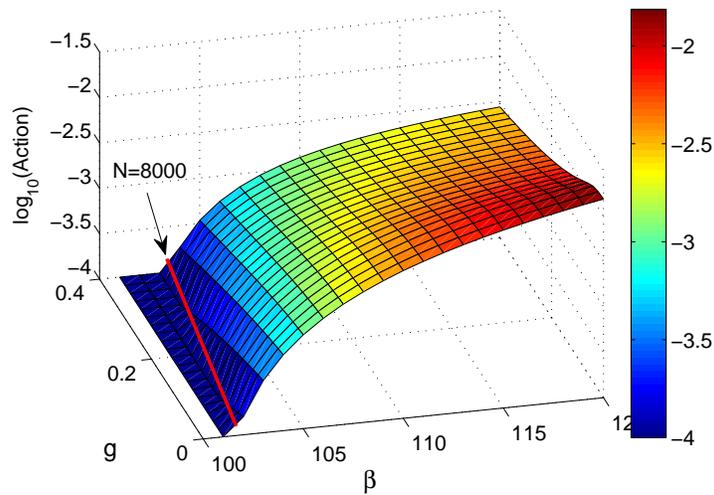} 
\end{center}
\caption{
{\bf Checking the threshold for quasi-stationarity.} A plot of the log of the action for Model 1 as we vary treatment, $g$, vs. the contact rate, $\beta$. In this case, the treatment frequency is $\nu=4$ year$^{-1}$. The red line indicates the threshold $1/N$ for  $N=8,000$ people. The action for $\beta=105$ year$^{-1}$ and $N=8,000$ people is greater than the threshold as we vary $g$. Therefore, we conclude these parameters would be sufficient for the model to exhibit extinction in the tail of the probability. As we increase the treatment fraction $g$ for $\beta=105$ year$^{-1}$, the action decreases towards the threshold line and a larger population would be necessary.
} 
\label{fig:thresholdnu4}
\end{figure}

To quantify where the system is quasi-stationary, we use the numerical approximation of the optimal path to evaluate Eq.~\eqref{eq:1DTaction} and construct a threshold graph to estimate the necessary parameters for the extinction to lie in the tail of the distribution. In Fig.~\ref{fig:thresholdnu4}, we show a threshold graph for the treatment model with frequency $\nu=4$ year$^{-1}$. For a disease with $\beta=105$, a population of $8,000$ is sufficiently to the right of the threshold for the exponent to be large for all $0<g<0.4$, meaning that the solution is quasi-stationary. 

The final step in finding the mean time to extinction is approximating the prefactor in Eq.~\eqref{eq:tau_ext}. Following the approach in \cite{Assaf2010}, we obtain\footnote{We use Eq.~(49) of \cite{Assaf2010} with $A_1=\frac{1}{R_0-1}$, which is the value that corresponds to our case.} 
\begin{equation}
B = \frac{1}{ (\beta-(\mu+\kappa)-\nu g)(R_0-1)} \sqrt{\frac{2\pi R_0 \nu g}{N (\mu+\kappa) \ln (1+\frac{\nu g}{\mu+\kappa} ) }}. 
\end{equation}
Note the dependence on the treatment parameters $g$ and $\nu$. 
 
\begin{figure}[!ht] 
\begin{center} 
\includegraphics[width=3.5in]{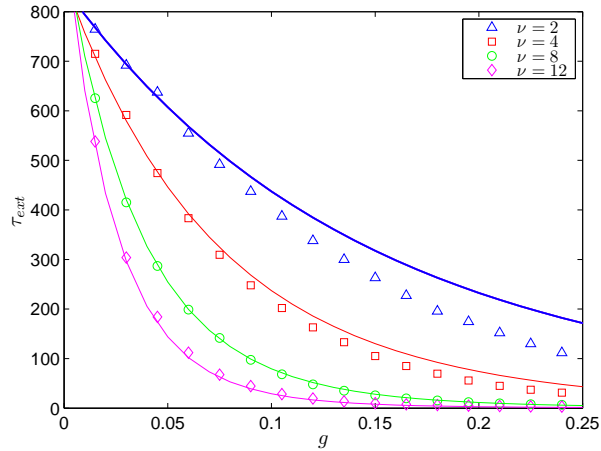} 
\end{center}
\caption{
{\bf The effectiveness of various treatment combinations for Model 1.} A plot of the fraction of infected treated, $g$, vs. the mean time to disease extinction, $\tau_{ext}$ years, for different treatment frequencies, $\nu$ year$^{-1}$. The results for the Monte Carlo simulations are averaged over 2,000 realizations and plotted as symbols. The curves of the same color show the approximation of the mean time to extinction by finding the action. The parameters are $N=8,000$ people and $\beta=105$ year$^{-1}$. Note the exponential decrease in the mean time to extinction as the treatment fraction is increased.
}
\label{fig:scaling10000} 
\end{figure}

To quantify the accuracy of the approximation to the mean time to extinction in Eq.~\eqref{eq:tau_ext}, with $\mathcal{S}$ up to $\mathcal{O}(N^{-1})$, we compare it to the average extinction time found by a Monte Carlo simulation as described in Gillespie \cite{Gill77}. In Figure \ref{fig:scaling10000}, the graph shows this comparison over a range of treatment percentages ($g$) and frequencies ($\nu$). The simulation uses a population of 10,000 and we averaged the results of 2,000 realizations. As expected, the mean time to extinction decreases as the treatment percentage and frequency increase. Note the excellent agreement for small $g$, for which the asymptotic approximation was derived.


\subsection*{Model 2}

\begin{figure}[!ht] 
\begin{center} 
\subfigure[]{\includegraphics[width=3.1in]{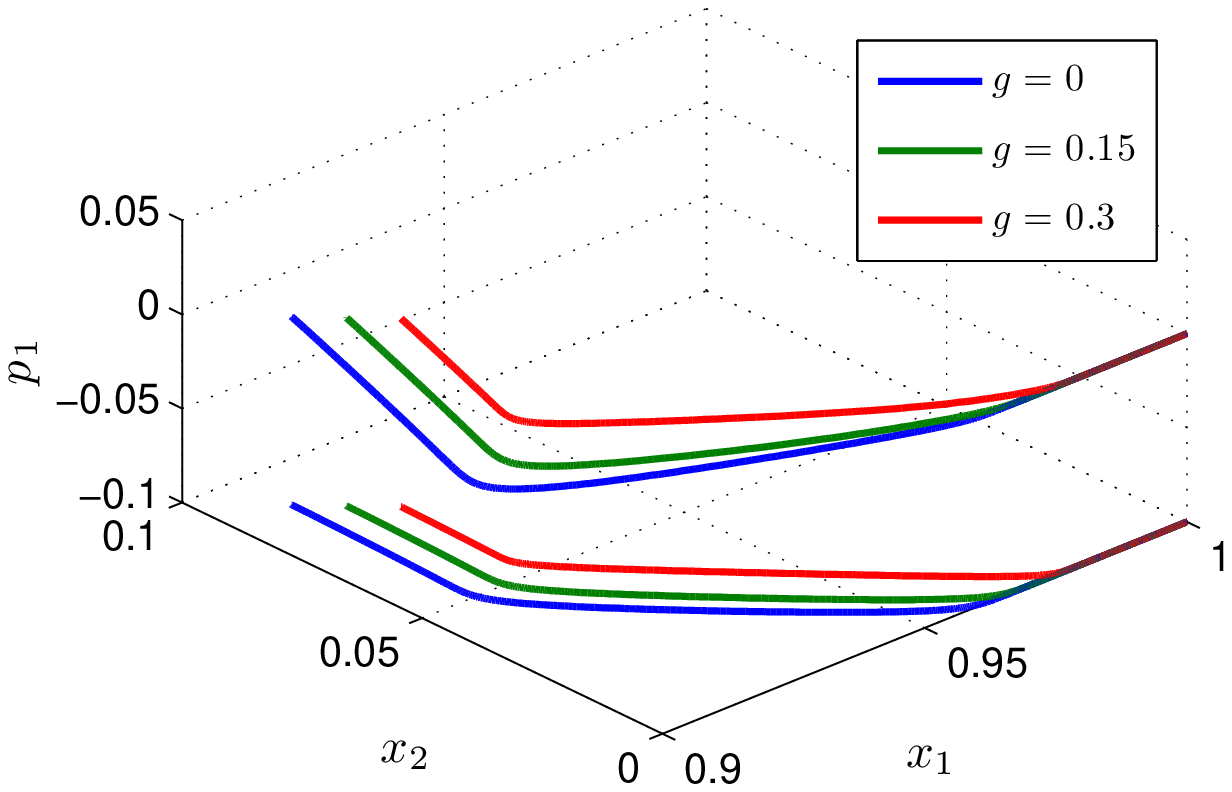}}
\subfigure[]{\includegraphics[width=3.1in]{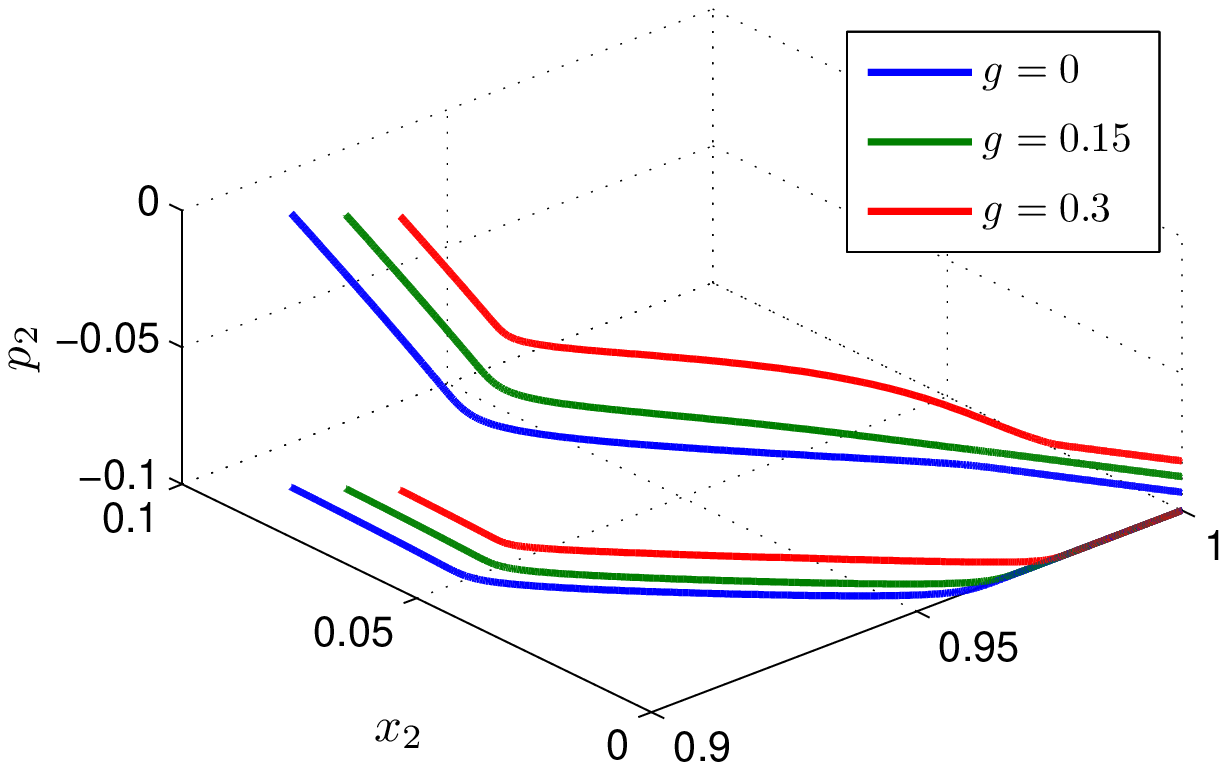}} 
\end{center}
\caption{
{\bf Graphs of the numerical approximations of optimal paths for Model 2.} In (a), we show the $(x_1,x_2,p_1)$ projection. In (b), we show the $(x_1,x_2,p_2)$ projection. The parameters are $\nu=4$ year$^{-1}$, $\beta=105$ and  $g=0, 0.15, 0.3$ for these examples. The arrow shows the direction of the path, connecting the endemic state to the extinct state. Below the curves are the projections of the $x_1$ and $x_2$ onto the respective momenta planes. 
}
\label{pathSIS2D} 
\end{figure}

\begin{figure}[!ht] 
\begin{center} 
\includegraphics[width=4in]{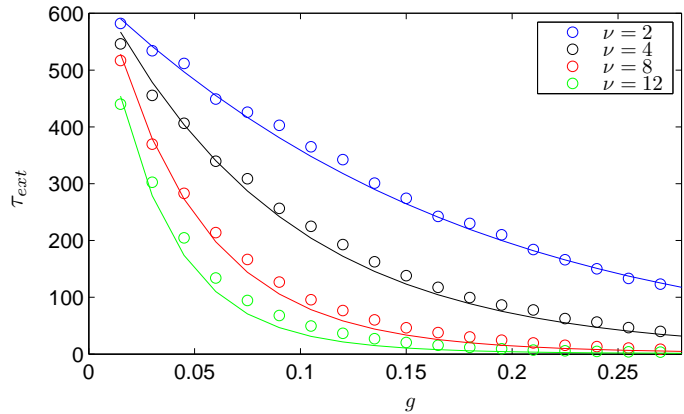} 
\end{center}
\caption{
{\bf The effectiveness of various treatment combinations for Model 2.} A plot of the fraction of infected vaccinated during each treatment, $g$, vs. the mean time to extinction, $\tau_{ext}$, for different treatment frequencies, $\nu$ year$^{-1}$. The average of 2,000 Monte Carlo simulations are shown with symbols. The curves of the same color show the numerical approximation of $\tau_{ext}$ using the action ${\cal S}$ and a constant prefactor.  For the parameters, we use $N=12,000$ people and $\beta=105$ year$^{-1}$.
} 
\label{fig:act2D} 
\end{figure}

The full SIS model has a Hamiltonian system in four dimensions and asymptotic approximations of the optimal path and action are not tractable. Therefore, we rely on numerical approximations. We show projections of the locations of several paths in phase space in Figure \ref{pathSIS2D}. The arrow shows the direction of the path, connecting the endemic state to the extinct state. Notice how the paths decrease in length as we increase the control parameter $g$. Note that the action from integrating along those paths also decreases with $g$. Using these paths, we can approximate the action and the mean time to extinction for a treatment schedule. We compare the theory to data found by Monte Carlo simulation for small $g$, as shown in Figure \ref{fig:act2D}. There is an exponential decrease in the mean time to extinction as we increase the treatment, agreeing with the theory.

\begin{figure}[!ht] 
\begin{center} 
\includegraphics[width=4in]{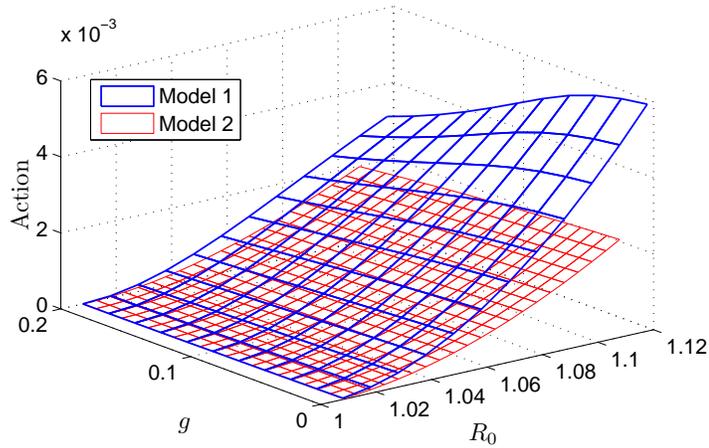} 
\end{center}
\caption{
{\bf A comparison of the action approximations for  Model 1 and Model 2.}  This plot shows the quantitative difference in the action approximation for Model 1 (blue) and Model 2 (red) as we vary $R_0$ and $g$. In this example, we use parameters $\beta = 105$ year$^{-1}$, $\nu=4$ year$^{-1}$, and $N=120,000$. 
}
\label{fig:compare_action} 
\end{figure}

We also comment on the differences in the constrained and unconstrained SIS models with treatment. In Figure \ref{fig:compare_action}, we compare the numerical approximations for actions of these models. We see that they agree for small $R_0$, but the action for the constrained model increases much faster as both $R_0$ and $g$ increase. This follows the result in \cite{sbd109}, where the action for the constrained SIS model was shown to have an exponential scaling law for $R_0 \gg 1$. The constrained treatment model also follows the exponential scaling. In addition, the theory can be used to avoid expensive simulations of long extinction times in large populations. The benefit of the full model is that it captures disease dynamics in a population with significant size fluctuations. The theory captures the rate of change in the mean time to extinction so that effectiveness in treatment schedules can be quantified.

\section*{Discussion}

In this paper, we quantified how treatment enhances the extinction of
epidemics using a stochastic, discrete-population framework. Specifically, we
based our study on a general formulation of an SIS model with treatment that is applied randomly in a Poisson fashion, accounting for the limited amount of resources. We used a WKB approximation to the master equation of the stochastic process to calculate the average time to extinction starting from the endemic state, as a function of the transmissibility of the disease and the strength and frequency of the treatment. We compared the extinction times obtained analytically and numerically from the WKB approximation with the values obtained from Monte Carlo simulations. 

In addition, we explored the significance of the quasi-stationarity assumption that is fundamental to the WKB approximation. The existence of a quasi-stationary distribution peaked at the endemic point produces a meta-stable state in which the population fluctuates in a neighborhood around the same endemic point. In contrast, the extinct state lies in the exponentially small tail of the distribution. When a quasi-stationary distribution exists, the extinction of a disease is a rare event, i.e. the mean time to extinction is exponentially long. As we show in the Supplementary Material, the time to extinction is indeed exponentially long when the disease-free point lies in the tail of the distribution. The occurrence of extinction as a rare event means that the fluctuations exhibited by the random population dynamics are much smaller than an effective activation barrier. If the fluctuations are not small compared to the barrier, then the extinction events are not necessarily in the tail of the distribution, and hence not a rare event. 

\begin{figure}[!ht] 
\begin{center} 
\includegraphics[width=4in]{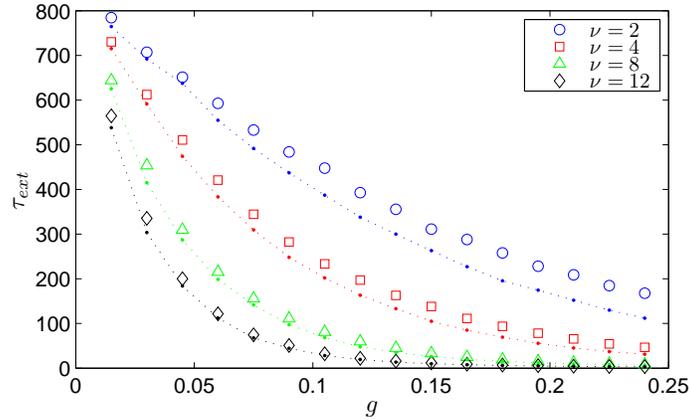} 
\end{center}
\caption{
{\bf A comparison of periodic and random treatment effectiveness.} For Model 1, a plot of the fraction of infected vaccinated during each treatment vs. the Monte Carlo simulated mean time to disease extinction for random (points connected by dotted lines) and periodic (symbols) treatment schedules. Results are shown for treatment frequencies, $\nu$ =2, 4, 8, and 12 year$^{-1}$ averaged over 2,000 realizations. The parameters are $N=8,000$ people and $\beta=105$ year$^{-1}$. Note that the random treatment schedule has average extinction times consistently lower than the periodic treatment schedule.
} 
\label{fig:periodic} 
\end{figure}

Deterministic models of treatment and/or vaccination are not accurate representations of the process in finite population realizations. A more realistic description is that, on average, treatment scheduling has a mean period or cycle, but is itself a random process. To quantify the difference between the deterministic and the stochastic descriptions, we compared the mean time to extinction for a strictly periodic and a Poisson-distributed treatment schedule obtained by averaging the Monte Carlo simulation results of many extinction events starting from the endemic state. We assume that a fraction $g$ of the infected population is treated at a frequency of $\nu$ times per year and immediately return to a susceptible state. In Figure \ref{fig:periodic}, simulations support evidence that the random schedule had a faster mean time to extinction over the range of frequencies. The reason for this may be that when the system is close to the extinct state, there is a benefit to having a number of treatment pulses in a short window of time; such a series of frequent vaccination pulses are possible in the Poisson vaccination scheduling but not in the deterministic one.

\begin{figure}[!ht] 
\begin{center} 
\includegraphics[width=4in]{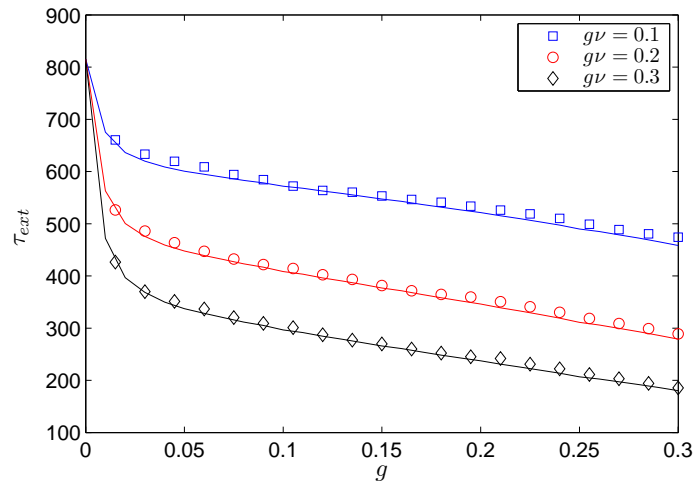} 
\end{center}
\caption{
{\bf  The effectiveness of various treatment combinations for a fixed treatment supply.} For Model 1, a plot of the mean time to extinction as we vary $g$ for a fixed treatment supply $g \nu =$constant. The symbols represent the Monte Carlo simulation results for $g\nu$ =0.1, 0.2, and 0.3 averaged over 10,000 realizations. The curves represent the direct numerical solution of the associated master equation. The parameters are $N=8,000$ people and $\beta=105$ year$^{-1}$. Note the decrease in mean time to extinction as the $g$ increases.
} 
\label{fig:fixedv} 
\end{figure}

The treatment program that we implement in our model has two degrees of freedom: the frequency $\nu$ and the fraction of infected individuals that are treated, $g$. On average, there are $\nu\cdot$(1 year) treatment pulses each year and at each one, a number $N g x_{2}$ of infected individuals are treated, where $x_2$ is the infected fraction at the moment each treatment pulse occurs. Supposing that there are a fixed number of treatment doses $N \nu g x_{2e}\cdot$(1 year)$=$constant that may be applied each year (here $x_{2e}$ is the fraction of the population that is infected at the endemic point). A natural question that arises is the following: Given a fixed number of total treatment doses, how are $\nu$ and $g$ chosen so that the time to disease extinction is minimized. In both of our SIS models, the fixed number of treatment doses translates into $\nu g = $ constant. Monte Carlo simulations of Model 1 show that the mean time to extinction decreases uniformly as $g$ increases, given a fixed $\nu g$ quantity (Fig.~\ref{fig:fixedv}). The drop is particularly sharp for $g\rightarrow 0$. This appears to be a consequence of the rounding down of $gX_2$ whenever a treatment pulse occurs (see Methods section). The treatment ceases to have an effect when there are less than $1/g$ infecteds; for very small $g$, the threshold $1/g$ is significant when compared to the number of infecteds at the endemic state. Thus, the treatment helps to bring the number of infected down to $1/g$, but not all the way to extinction. This issue does not appear if one instead chooses to round $gX_2$ to the next-highest integer (results not shown). With this alternative method of rounding, the time to extinction actually has a sharp decrease as $g\rightarrow 0$. Monte Carlo simulations of Model 2 corroborate this finding. Thus, given a fixed number of resources, our stochastic simulations demonstrate that in order to eliminate infectious diseases, it is better to increase the pool of individuals reached by the treatment, rather than increase its frequency.

In conclusion, we have described a method to quantify the effectiveness of a random treatment program. We find that increasing the magnitude and frequency of randomly scheduled treatments provide an exponential decrease in average extinction times. We have presented evidence that supports how larger campaigns applied less frequently are the most effective in facilitating disease eradication. Several assumptions in the model clarify the accuracy of the analytic approximation to the mean time to extinction, but its exponential rate of decrease as we increase the intervention is consistent with simulations throughout our analysis as populations get very large.  The techniques considered here can be easily generalized to other diseases, such as those that include seasonality or population structure. Future work in this area could provide a more targeted control strategy that would be robust in fluctuating environments as well as more efficient and economical disease eradication.  
 

\clearpage

\section*{Supplementary Material}
 
\subsection*{Quasi-stationarity}

In this section, we illustrate the central importance of the quasi-stationarity assumption for the accuracy of the WKB expression for the mean-extinction time. For this, we numerically compute solutions to the master equation for the constrained SIS model without treatment (Model 1 with $g=0$), and determine when the WKB condition $N {\cal S}_{opt} \gg 1$ holds and when it does not by comparing the outcome with the WKB results. 

 In the normalized constrained variables where $x_1 +x_2 = 1$ (i.e., no fluctuations in the total population), the Hamiltonian takes the form:
 \begin{equation} 
\begin{array}{l} 
H(x_2,p_2) = \beta x_2 (1-x_2) (e^{p_2}-1)+ (\mu+\kappa) x_2 (e^{-p_2}-1). 
\end{array} 
\end{equation} 
The Hamiltonian system under the constraint $H(x_2,p_2)=0$ has the steady states 
 \begin{equation} 
 (x_{2e},p_{2e}) = \left(1 - \frac{1}{R_0}- \frac{\nu g}{\beta},0 \right)
\end{equation} 
 and $(x_{20},p_{20}) = (0,-\ln{R_0})$, which denote the endemic and extinct states, respectively. The system also possesses a third steady state $(x_{2m}, p_{2m}) = (0,0)$. The attracting states $(x_{2e},p_{2e})$ and $(x_{2m}, p_{2m})$ correspond to the zero fluctuation states that exist in the mean field equation (deterministic system). Since there is a nonzero probability current at the extinct state, $(x_{20},p_{20})$ is a new state created by the noise in the system. We also note that for finite populations
noise is not generally known due to the random interactions of individuals. However, in this model, the noise-free extinct state is accessible if the noise is known to be Gaussian. 

The most probable path from the endemic to the extinct point is the heteroclinic trajectory connecting the fixed point $(x_{2e},p_{2e})$ with the fluctuational extinction point $(x_{20},p_{20})$. The optimal path is given by $p_2(x_2)=-\ln(R_0(1-x_2))$. Therefore, the action from the endemic state to a point $x_2$ along the optimal path up to the zeroth order of $N$ is
\begin{equation} \label{eq:1Daction} 
 {\cal S}(x_2) = \int_{\frac{R_0-1}{R_0}}^{x_2} -\ln(R_0(1-x'_2)) \, dx'_2. 
\end{equation} 
In particular, the action from the endemic state to the extinct state along the optimal path is 
\begin{equation} 
 {\cal S}_{opt} = {\cal S}(0) = -1 + \ln(R_0) + \frac{1}{R_0}. 
\end{equation} 
Since $R_0 > 1$, for sufficiently large $N$ we have $N {\cal S}_{opt} \gg 1$, ensuring the extinct point lies in the tail of the probability distribution, where its value $\rho(0) = A e^{-N{\cal S}_{opt}}$ is exponentially small.

\begin{figure} 
\begin{center}
 \includegraphics[height=70mm]{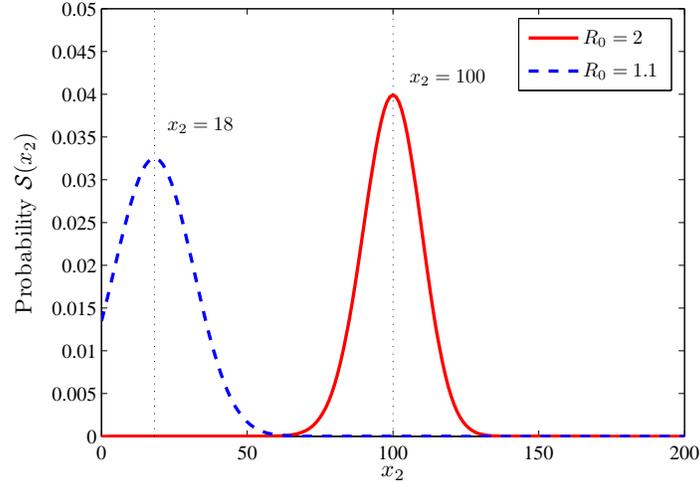} 
 \end{center}
\caption{
{\bf Quasi-stationarity depicted through probability distributions.} Graphs of the WKB approximation of the SIS probability distributions using Eq.~(\ref{eq:1Daction}) for $N=200$. We show the case of $R_0=2$, for which extinction is in the tail of the distribution. Conversely, extinction has a significant probability in the case of $R_0=1.1$. Note the height of the curve for $x_2=0$. The dotted vertical lines show the location of the endemic state in $x_2$ for each case. 
} 	
\label{fig:problimit}	
\end{figure}

In Figure \ref{fig:problimit}, the WKB approximations to the quasi-stationary distribution are shown for the cases of $R_0=2$ and $R_0=1.1$. The value of the distribution at zero shows that the extinction point is definitely not in the tail of the distribution for $R_0=1.1$ and hence does not constitute a rare event. In contrast, for $R_0=2$, the extinct state is in the tail of the distribution and hence we expect the WKB result to be accurate.

\begin{figure}[!ht] 
\begin{center} 
\subfigure[]{\includegraphics[height=50mm]{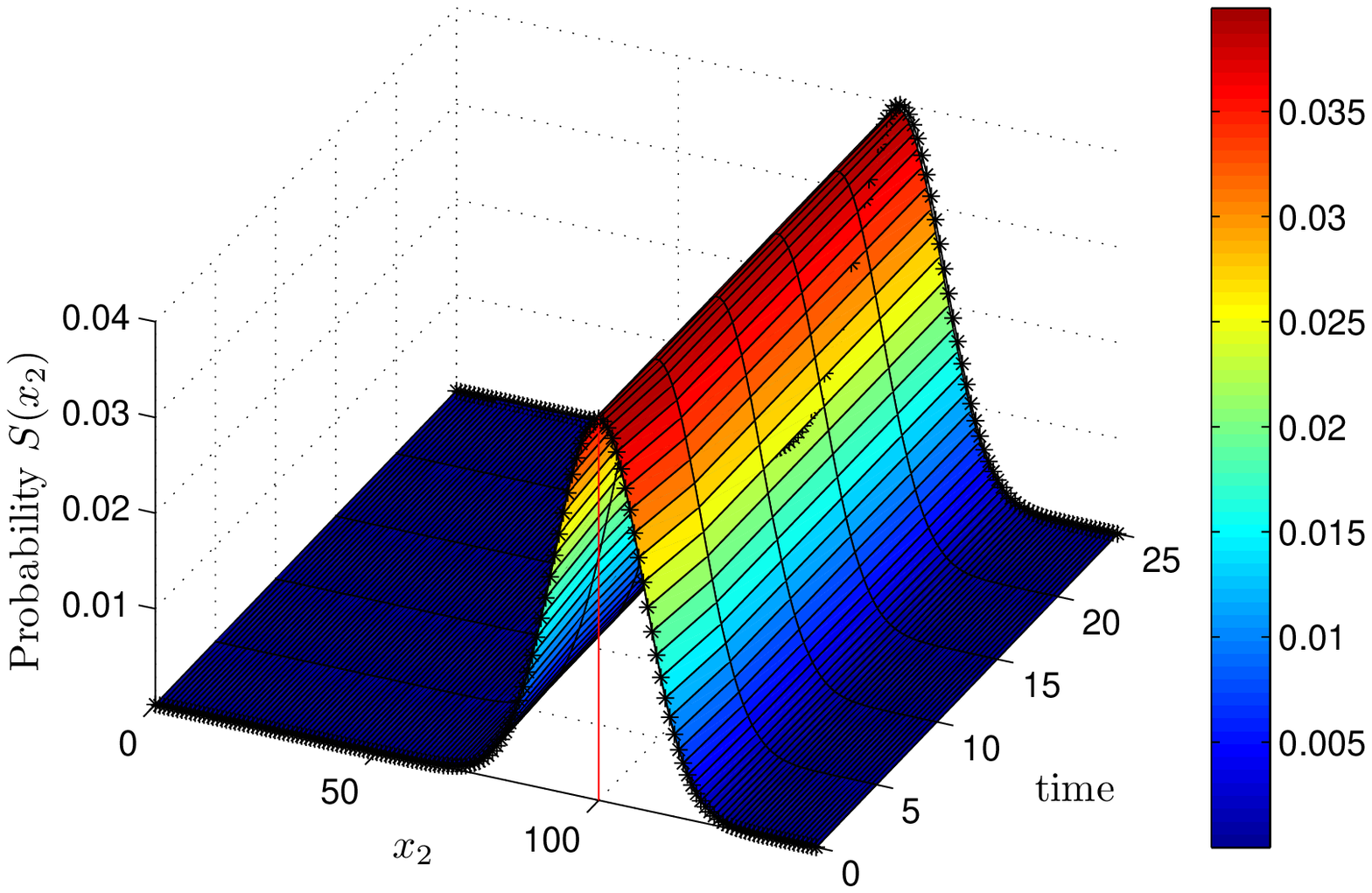} } 
\subfigure[]{\includegraphics[height=50mm]{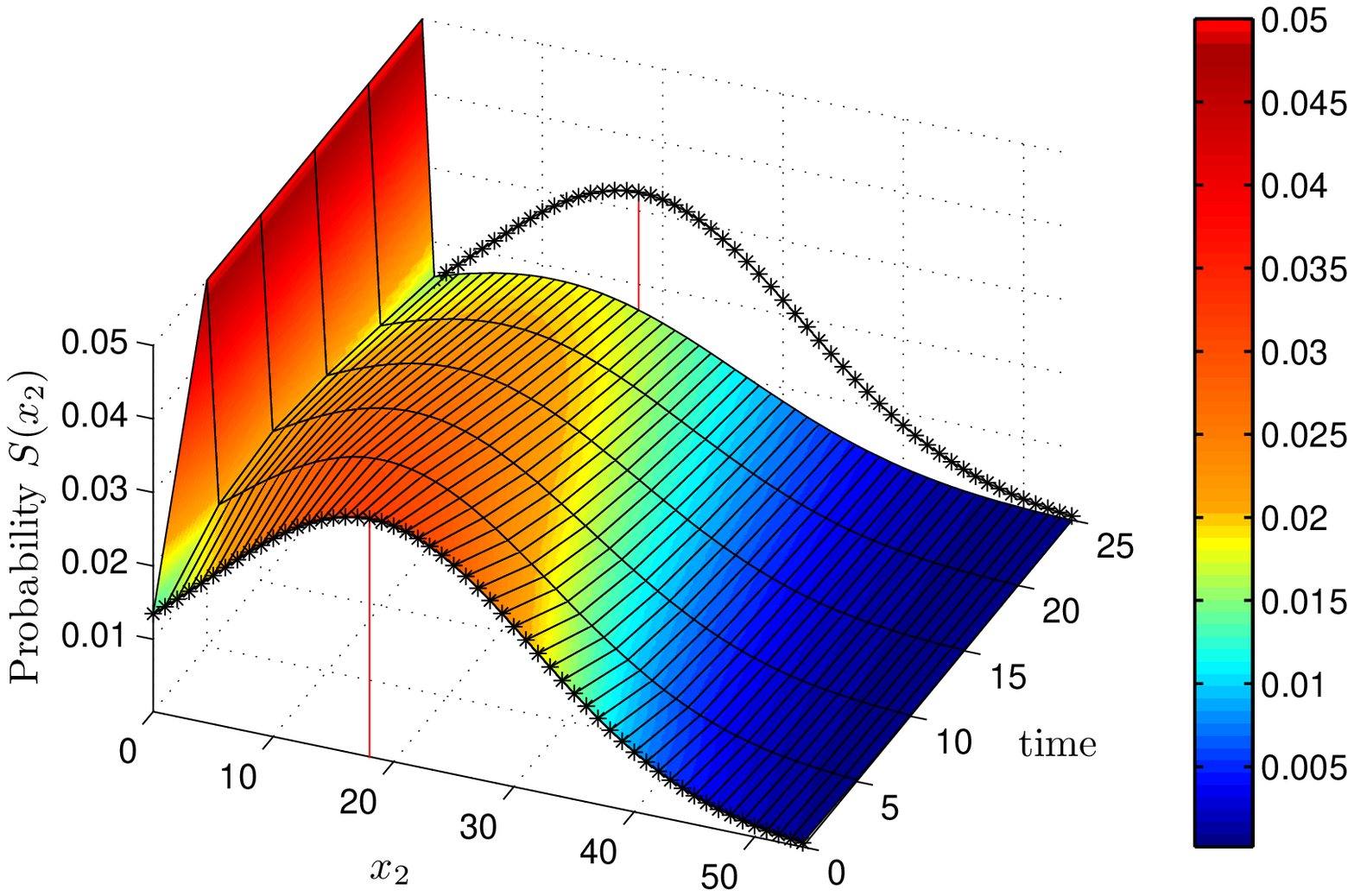} }
 \end{center}
\caption{
{\bf The drift of probability distributions for systems without quasi-stationarity.} A plot of the solution of the 1D master equation in Eq.~\eqref{master:1DSIS} with $g=0$ over time using the distribution from the WKB approximation, Eq.~(\ref{eq:1Daction}), as initial condition. For $R_0=2$ (left), the extinct state lies in the tail of the distribution and a quasi-stationary distribution exists. Extinction occurs only over exponentially long times. For $R_0=1.1$ (right) the endemic state is close to the absorbing boundary and extinction is not a rare event. The absorption of this distribution into the boundary is apparent.
}
\label{fig:action}
\end{figure}

Because the disease free equilibrium is always an absorbing boundary in the one-dimensional case, the system decays to the extinct state in the long term. If a quasi-stationary distribution exists, the complete decay happens for exponentially long times; otherwise, it occurs on a much shorter time-scale. To illustrate this phenomenon, we show the numerical solution of the master equation in Eq.~(\ref{master:1DSIS}) over time in Figure \ref{fig:action}. The initial probability distribution at $t=0$ is set to the WKB approximation of the SIS probability distributions using Eq.~(\ref{eq:1Daction}). The absorption into the disease-free state is apparent in the $R_0=1.1$ case, but completely imperceptible for $R_0=2$ over the time-scale shown.

 \section*{Acknowledgments} 
 We gratefully acknowledge support from the Office of Naval Research . L.B., L.M., and I.B.S. are supported by the National Institute of General Medical Sciences Award No. R01GM090204. The content is solely the responsibility of the authors and does not necessarily represent the official views of the National Institute of General Medical Sciences or the National Institutes of Health. B.L. is currently an NRC Postdoctoral Fellow. 

\bibliography{sources_vac,Refs_IBS}

\clearpage

\section*{Figure Legends}


 \end{document}